\documentclass[10pt,aps,pra,twocolumn,superscriptaddress]{revtex4-2}

\usepackage{graphicx}  
\usepackage{subfigure}
\usepackage{multirow}

\usepackage{xcolor}
\usepackage{fancyhdr}
\usepackage{longtable}
\usepackage{parskip}
\usepackage[T1]{fontenc}
\usepackage{dcolumn}   
\usepackage{booktabs}
\usepackage{bm}        
\usepackage{amsfonts}  
\usepackage{amsmath}   
\usepackage{amssymb}   
\usepackage{gensymb}

\definecolor{verdemilitar}{RGB}{93, 101, 50}

\newcommand{\pwisein}{\left\{ \begin{array}{ll}}
\newcommand{\pwiseout}{\end{array}\right.}
\newcommand{\ket}[1]{\left| #1 \right\rangle}

\makeatletter

  {\par\begingroup\small\itshape\color{red}\noindent[SPW:\ }%
  {]\par\endgroup\noindent\ignorespacesafterend}
  {\par\begingroup\color{blue}}%
  {\par\endgroup\noindent\ignorespacesafterend}
\makeatother

\date{July 2024}
\begin{document}

\title{Hyperentanglement in intermediate phase-space planes: benchmarking a
photon-pair source for quantum key distribution}

\author{Murilo H. Magiotto}
\affiliation { Instituto de F\'isica, Universidade Federal de Goiás, 74690-900, Goiânia, Goiás, Brazil}
\author{Millena B. Novaes}
\affiliation { Instituto de F\'isica, Universidade Federal de Goiás, 74690-900, Goiânia, Goiás, Brazil}        \author{Wesley B. Cardoso} 
\affiliation { Instituto de F\'isica, Universidade Federal de Goiás, 74690-900, Goiânia, Goiás, Brazil}
\author{Ardiley T. Avelar}
\affiliation { Instituto de F\'isica, Universidade Federal de Goiás, 74690-900, Goiânia, Goiás, Brazil}
\author{Paulo H. Souto Ribeiro}
\affiliation{Departamento de F\'{i}sica, Universidade Federal de Santa Catarina, 88040-900, Florian\'{o}polis, SC, Brazil}
\author{Stephen P. Walborn}
\affiliation { Departamento de Física, Universidad de Concepción, 160-C, Concepción, Chile}
\affiliation { Millennium Institute for Research in Optics, Universidad de Concepción, 160-C, Concepción, Chile}
\author{Guilherme L. Zanin}
\affiliation { Instituto de F\'isica, Universidade Federal de Goiás, 74690-900, Goiânia, Goiás, Brazil}
\author{Rafael M. Gomes}
\email{rafaelgomes@ufg.br}
\affiliation { Instituto de F\'isica, Universidade Federal de Goiás, 74690-900, Goiânia, Goiás, Brazil}

\date{\today}

\begin{abstract}  
We characterize a source of photon pairs simultaneously entangled in polarization and in
the transverse spatial degree of freedom, with the spatial measurements performed in
intermediate planes of the fractional Fourier transform. The source consists of two crossed
type-I crystals in a non-collinear geometry, such that the transverse variables carry
information about the crystal of origin. Detection in the crystal plane reveals this information
completely and destroys the polarization entanglement, so that a scheme using the
near and far fields as conjugate spatial bases cannot support a parallel polarization channel. In
intermediate planes both spatial bases combine position and momentum, such that the which-crystal
information is partially erased, and the polarization entanglement survives in all four
combinations of spatial bases. Applying
Bell and Einstein-Podolsky-Rosen criteria to the same photon pairs certifies both degrees of
freedom simultaneously. We evaluate quantum bit error rates and asymptotic secret key rates in
each degree of freedom, and find that the source delivers two simultaneously usable channels
per photon pair.
\end{abstract}

\maketitle 

\section{Introduction}

Hyperentanglement, the simultaneous entanglement of two or more degrees of freedom of the same
photon pair \cite{Kwiat97}, is a well established resource for quantum information processing
\cite{Deng17}. Photon pairs hyperentangled in polarization, momentum,
orbital angular momentum and energy-time have been produced and characterized
\cite{Barreiro05,Barbieri05,Cinelli05,Vallone09}. Hyperentanglement has enabled complete Bell-state
analysis with linear optics \cite{Kwiat98,Walborn03,Schuck06,Barbieri07,Wei07}, dense coding beyond
the linear-optics channel capacity \cite{Barreiro08}, superdense teleportation \cite{Graham15},
entanglement purification \cite{Simon02,Hu21}, enhanced tests of nonlocality
\cite{Cinelli05,Barbieri06}, two-photon cluster states for one-way quantum computation
\cite{Vallone07,Ceccarelli09}, the determination of entanglement with a single measurement
\cite{Walborn06Nature}, enhanced precision in quantum metrology \cite{Walborn2018}, and it
enlarges the accessible Hilbert space without increasing the number of photons
\cite{malik2026highdimensionalquantumphotonicsroadmap}. Photon pairs produced by
spontaneous parametric down-conversion (SPDC) are naturally entangled in several degrees of freedom
\cite{Kwiat97}. Among these, polarization and the transverse spatial variables
are especially attractive, since they can be manipulated independently with linear optics, and
together they combine a discrete variable with a continuous one.

In most applications the degrees of freedom are treated as independent, so that the state is taken
to be a product of entangled states, one in each degree of freedom. Whether a given source delivers
this product depends on how the photons are measured as well as on how they are produced. The
two-crystal type-I source \cite{Kwiat99}, one of the most widely used sources of polarization
entanglement, relies on the indistinguishability of the pairs produced in one crystal or the other.
In the usual non-collinear geometry the pairs produced in the two crystals are displaced laterally
with respect to one another, so that the transverse variables carry information about the crystal of
origin \cite{Caetano2003OC,Altepeter2005,Rangarajan2009}. Detection in the crystal plane reads this
information out and leaves no polarization entanglement, while detection in a basis that does not
resolve the displacement eliminates it, as in a quantum eraser \cite{Scully82,Walborn2002eraser}.
The coupling is largely avoided when the spatial variables are measured in the far field alone \cite{Barbieri05}.
However, tasks that require complementary spatial measurements, such as the certification of
spatial entanglement and Einstein-Podolsky-Rosen (EPR) steering \cite{Reid89,Howell04} or
entanglement-based cryptography, need two conjugate spatial bases. The usual choice of the near and
far fields then loses the polarization entanglement whenever the near-field basis is measured.

Here we characterize a photon-pair source hyperentangled in polarization and in the transverse
spatial degree of freedom, with the spatial measurements performed in intermediate planes of the
fractional Fourier transform (FRFT) \cite{Tasca08,PhysRevA.79.033801,tasca09b}. These planes
provide two conjugate spatial bases, each of which combines position and momentum. We show that
the which-crystal information is then strongly suppressed in both bases, such that the
polarization entanglement survives in all four combinations of spatial bases, and the two degrees of
freedom are available in parallel.

To quantify the resource we use the figures of merit of entanglement-based quantum key distribution
(QKD) \cite{Bennett84,Ekert91,Gisin02,Xu20}. QKD provides well-defined operational measures, the
quantum bit error rate and the secret key rate, and it is known that encoding in higher dimensions
or in the spatial variables increases the key per detected photon
\cite{Groblacher06,Walborn06PRL,Almeida2005,Leach2012}. A source that is entangled in each degree of
freedom, in every basis used, provides a key in each. If, in addition, the state factorizes between
the two degrees of freedom, the source provides two independent channels. We apply Bell and EPR
criteria to the same photon pairs, evaluate the error rates and asymptotic secret key rates in each
degree of freedom, and use the joint error statistics to test the factorization. We find that the
source delivers two simultaneously usable channels per photon pair, where
the key rates serve as a benchmark of the source. The same requirements, entanglement in conjugate
bases of each degree of freedom and factorization between them, can apply to other protocols that use
the two degrees of freedom in parallel.

\section{Hyperentangled States}
\label{sec:hyper}

SPDC naturally generates photon pairs entangled in several degrees of freedom, making it one
of the most versatile resources available for experimental quantum information science
\cite{Kwiat95,Bouwmeester97,Jennewein00,Pan98,Pan00,Knill01,Howell04,Walborn10,Walborn11,Deng17}.
Among the available degrees of freedom, polarization and transverse spatial variables are
especially attractive because they can be easily manipulated and controlled independently. Their
coexistence provides a natural physical platform for realizing hybrid quantum communication
protocols based on hyperentanglement.

\subsection{SPDC source in near and far field}
\label{sec:source}
For now we consider only the spatial degrees of freedom.
We label the two photons $A$ and $B$, and denote their transverse positions by
$\bm{\rho}_{A,B}$ and their transverse wavevectors by $\mathbf{q}_{A,B}$. For simplicity we
take these to be dimensionless, which is achieved by dividing or multiplying by a
characteristic length $\gamma$ defined below, so that
$[\rho_{A},q_{A}]=[\rho_{B},q_{B}]=i$. The global variables are
\begin{equation}
\bm{\rho}_{\pm}=\frac{\bm{\rho}_{A}\pm\bm{\rho}_{B}}{\sqrt{2}},
\qquad
\mathbf{q}_{\pm}=\frac{\mathbf{q}_{A}\pm\mathbf{q}_{B}}{\sqrt{2}},
\label{globals}
\end{equation}
which form two canonically conjugate pairs, $[\rho_{+},q_{+}]=[\rho_{-},q_{-}]=i$, while
$[\rho_{+},q_{-}]=[\rho_{-},q_{+}]=0$. In the Gaussian approximation the two-photon
amplitude generated in a single crystal is \cite{Walborn10,Schneeloch_2016}
\begin{equation}
\psi(\mathbf{q}_{+},\mathbf{q}_{-}) \propto
\exp\!\left[-\frac{\mathbf{q}_{+}^{2}}{4\sigma_{+}^{2}}\right]
\exp\!\left[-\frac{\mathbf{q}_{-}^{2}}{4\sigma_{-}^{2}}\right],
\label{gauss}
\end{equation}
where $\sigma_{+}$ is fixed by the angular spectrum of the pump and $\sigma_{-}$ by the
phase-matching function, and $\sigma_{+}\ll\sigma_{-}$ for typical parameters. The state is
narrow in $\mathbf{q}_{+}$ and, correspondingly, narrow in $\bm{\rho}_{-}$, so the photon
pair is correlated in position and anticorrelated in momentum.

It will be convenient to introduce the variance of a quadrature
rotated by an angle $\zeta$ in the phase space of a mode whose momentum spread is $\sigma$,
\begin{equation}
\Lambda_{\zeta}(\sigma) \equiv \frac{\cos^{2}\zeta}{4\sigma^{2}} + \sin^{2}\zeta\;\sigma^{2} .
\label{Lam}
\end{equation}
The variances of the global variables are then
\begin{equation}
\Delta^{2}\bm{\rho}_{\pm}=\Lambda_{0}(\sigma_{\pm})=\frac{1}{4\sigma_{\pm}^{2}},
\qquad
\Delta^{2}\mathbf{q}_{\pm}=\Lambda_{\pi/2}(\sigma_{\pm})=\sigma_{\pm}^{2} ,
\label{widths}
\end{equation}
each sector saturating $\Delta^{2}\rho_{\pm}\Delta^{2}q_{\pm}=1/4$. Since the state
factorizes between the two sectors and each factor is even, all cross moments vanish. In
particular $\langle\bm{\rho}_{-}\mathbf{q}_{+}\rangle=0$, so the variances of the two narrow
variables simply add wherever they appear together.

Imaging the crystal onto the detectors measures $\bm{\rho}_{A}$ and $\bm{\rho}_{B}$, so the
spatial correlation appears in the difference of the detector coordinates, with variance
$\Delta^{2}\bm{\rho}_{-}$. A Fourier transform gives access to $\mathbf{q}_{A}$ and
$\mathbf{q}_{B}$, so the correlation appears in the sum, with variance
$\Delta^{2}\mathbf{q}_{+}$. This is the basic asymmetry of standard SPDC: each
conjugate basis is governed by one source parameter. Bringing the
two correlation widths to a common scale requires $2\sigma_{+}\sigma_{-}=1$, a condition
that fixes $\gamma$, and hence the focal lengths of the detection optics, in terms of the
pump and the crystal. For a typical source and a convenient choice of optics the product is
one or two orders of magnitude from unity.  The
asymmetry holds for a perfect source, one in which every coupling between the spatial and
polarization degrees of freedom has been minimized by--say--using a Sagnac geometry, a periodically
poled crystal, or a crystal thin enough that propagation effects are negligible.
\subsection{Intermediate planes}
\label{sec:intermediate}

Spatial correlations can also be observed between the near and far fields, in the planes of
the fractional Fourier transform (FRFT) \cite{Tasca08,PhysRevA.79.033801,tasca09b}. In
dimensionless coordinates an FRFT of order $\alpha$ is a rotation of phase space, and the
transverse coordinate recorded in the corresponding plane is
\begin{equation}
\bm{u}^{(\alpha)}=\bm{\rho}\cos\alpha+\mathbf{q}\sin\alpha ,
\label{frft1}
\end{equation}
with $\alpha=0$ the crystal plane and $\alpha=\pi/2$ the far field. Transforms of any order can
be implemented with one or a few lenses \cite{pellat-finet94,ozaktas01}. We use the two
intermediate bases
\begin{equation}
\bm{r}\equiv\bm{u}^{(\pi/4)}=\frac{\bm{\rho}+\mathbf{q}}{\sqrt{2}},
\qquad
\bm{s}\equiv\bm{u}^{(3\pi/4)}=\frac{-\bm{\rho}+\mathbf{q}}{\sqrt{2}},
\label{rsdef}
\end{equation}
whose orders differ by $\pi/2$, so that $r$ and $s$ are mutually unbiased in the same sense as
position and momentum. FRFT bases of this kind have been used to detect continuous-variable
entanglement \cite{Tasca08,PhysRevA.94.012303} and to simulate post-quantum correlations
\cite{tasca09b}.

The spatial correlations are governed by the sum and difference of the detected coordinates,
$\bm{u}_{\pm}=(\bm{u}_{A}\pm\bm{u}_{B})/\sqrt{2}$. When the two photons are transformed with
different orders, each of these combinations mixes a global position with the momentum of the
other global sector, $\bm{\rho}_{\pm}$ with $\mathbf{q}_{\mp}$, which a common order applied to
both photons never does \cite{companion}. This is what allows the two narrow source variables,
$\bm{\rho}_{-}$ and $\mathbf{q}_{+}$, to appear in the same measured combination. For the four
configurations used here, Eqs.~(\ref{frft1}) and (\ref{rsdef}) give directly
\begin{equation}
\bm{u}_{\pm}=\frac{\epsilon\,\bm{\rho}_{\nu}+\mathbf{q}_{\pm}}{\sqrt{2}},
\qquad
\nu=\begin{cases}\mp & \text{opposite bases,}\\[2pt] \pm & \text{same basis,}\end{cases}
\label{intmap}
\end{equation}
with $\epsilon=+1$ when photon $A$ is measured in $r$ and $\epsilon=-1$ when it is measured in
$s$. The two global sectors of the state of Eq.~(\ref{gauss}) are independent, so that
\begin{equation}
\Delta^{2}\bm{u}_{\pm}
=\tfrac{1}{2}\left[\Delta^{2}\mathbf{q}_{\pm}+\Delta^{2}\bm{\rho}_{\nu}\right],
\label{intvar}
\end{equation}
the average of a far-field and a near-field width in every case. In opposite bases the sum
$\bm{u}_{+}$ contains both narrow variables, $\bm{\rho}_{-}$ and $\mathbf{q}_{+}$, while in the
same basis it contains $\bm{\rho}_{+}$ and $\mathbf{q}_{+}$, of which only the second is narrow.

Since the state is real and factorizes between the two sectors, the cross moment
$\langle\bm{u}_{+}\bm{u}_{-}\rangle$ vanishes in all four configurations. The two detectors then
have the same marginal variance, $(\Delta^{2}\bm{u}_{+}+\Delta^{2}\bm{u}_{-})/2$, and the
conditional variance of one detected coordinate given the other is
\begin{equation}
\Delta^{2}\!\left(\bm{u}_{A}|\bm{u}_{B}\right)
=\frac{2\,\Delta^{2}\bm{u}_{+}\,\Delta^{2}\bm{u}_{-}}
      {\Delta^{2}\bm{u}_{+}+\Delta^{2}\bm{u}_{-}} .
\label{condvar}
\end{equation}
For a strongly correlated source, $2\sigma_{+}\sigma_{-}\gg1$, the sum is the narrow combination
in all four configurations, and Eqs.~(\ref{widths}), (\ref{intvar}) and (\ref{condvar}) give
\begin{align}
\Delta^{2}\!\left(\bm{u}_{A}|\bm{u}_{B}\right)_{\rm opp}
&\simeq\sigma_{+}^{2}+\frac{1}{4\sigma_{-}^{2}},
\label{condopp}\\
\Delta^{2}\!\left(\bm{u}_{A}|\bm{u}_{B}\right)_{\rm same}
&\simeq\sigma_{+}^{2}+\frac{1}{4\sigma_{+}^{2}}.
\label{condsame}
\end{align}
The opposite-basis value is the mean of the far-field and near-field conditional variances,
$2\sigma_{+}^{2}$ and $1/2\sigma_{-}^{2}$. In the same basis the near-field contribution comes
instead from the broad variable $\bm{\rho}_{+}$, whose width is set by the pump. The photons
remain correlated, since Eq.~(\ref{condsame}) is still much smaller than the marginal variance,
but the conditional variance is larger by a factor of approximately $1+1/4\sigma_{+}^{4}$. The product of the two same-basis conditional variances is then larger than the bound of
Eq.~(\ref{EPR}), and only the opposite-basis configurations can show steering.

\subsection{Spatial-Polarization Coupling}
\label{sec:coupling}

The source used here consists of two type-I crystals of thickness $\ell$ with orthogonal optic
axes, placed end to end, so that the state is
\begin{equation}
|\Psi\rangle = |\mathbf{e}_{0}\mathbf{e}_{0}\rangle\,|\psi_{1}\rangle
             + |\mathbf{e}_{1}\mathbf{e}_{1}\rangle\,|\psi_{2}\rangle ,
\label{I}
\end{equation}
where $\ket{\psi_{1}}$ and $\ket{\psi_{2}}$ are the spatial states produced in the first and
second crystals, referred to a common plane. The two crystals occupy
different longitudinal positions and the pairs they produce carry which-crystal information in
their transverse variables
\cite{Kwiat99,Nambu02,Caetano2003OC,Caetano2003PRA,Rangarajan2009,Osorio2007}.

In the non-collinear geometry used here the down-converted photons are collected at angles
$\pm\theta_{e}$ to the pump in the horizontal ($xz$) plane, corresponding to $\pm\theta_{i}$ inside the
crystal, with $\sin\theta_{i}=\sin\theta_{e}/n$, and the phase mismatch is dominated by a term linear in the relative transverse wave
vector \cite{MolinaTerriza2005,Torres2005}. A
pair generated at depth $z$ in a crystal separates laterally as it leaves the crystal, so the
relative positions of the pairs from one crystal are uniformly distributed over a width
$W=\sqrt{2}\,\ell\sin\theta_{i}$ in $\rho_{-,x}$. The pairs from the first crystal are born a
distance $L$ earlier along the pump, where $L$ is the separation between the centers of the
two crystals, and are displaced with respect to those of the second crystal,
\begin{equation}
\psi_{1}(\bm{\rho}_{+},\bm{\rho}_{-})\simeq\psi_{2}(\bm{\rho}_{+},\bm{\rho}_{-}-d_{-}\hat{\mathbf{x}}),
\label{shiftx}
\end{equation}
with $d_{-}=\sqrt{2}\,[\ell\sin\theta_{i}+(L-\ell)\sin\theta_{e}]$, since the photons travel a
distance $\ell$ inside the crystals and $L-\ell$ in the gap between them. For crystals in contact
$L=\ell$ and $d_{-}=W$.
For $\ell=1$~mm and $\theta_{e}\simeq3^{\circ}$ this gives $d_{-}\simeq52~\mu$m, or
$d_{-}/\gamma\simeq0.017$ in the dimensionless units of Sec.~\ref{sec:intermediate}. Birefringent
walk-off adds small translations of the pair center $\bm{\rho}_{+}$, and the additional
propagation of the pairs from the first crystal adds a phase that dominates in the collinear
geometry but is negligible here \cite{companion}.

A pair of detection apertures of finite size records an incoherent sum of the pure polarization states obtained
at each pair of detected coordinates. The result has populations in
$\ket{\mathbf{e}_{0}\mathbf{e}_{0}}$ and $\ket{\mathbf{e}_{1}\mathbf{e}_{1}}$ and a single
coherence, and its concurrence \cite{Wootters98} is
\begin{equation}
\mathcal{C}=\frac{2|I_{12}|}{I_{11}+I_{22}},
\qquad
I_{jk}=\int d\bm{u}_{A}\,d\bm{u}_{B}\;M_{A}M_{B}\,\Psi_{j}\Psi_{k}^{*},
\label{conc}
\end{equation}
where $\Psi_{j}(\bm{u}_{A},\bm{u}_{B})$ is the amplitude of state $j$  at the detection plane and
$M_{A,B}$ are the transmission profiles of the two detectors. The coherence $I_{12}$ is the
overlap of the spatial states of the two crystals, restricted to the collected region. In the
crystal plane (near field) the detectors measure the positions $\bm{\rho}_{A}$ and $\bm{\rho}_{B}$
themselves. 
The two distributions of $\rho_{-,x}$ in Eq.~(\ref{shiftx}) have width $W$ and are offset by
$d_{-}$, so they overlap only if $d_{-}<W$.  For crystals in contact $d_{-}=W$, and any gap makes $d_{-}$ larger, so that no pair of detector positions in the near field receives light from both crystals and the
coherence vanishes. The measurement thus reveals the crystal of origin and leaves the polarization in a
mixture of $\ket{\mathbf{e}_{0}\mathbf{e}_{0}}$ and $\ket{\mathbf{e}_{1}\mathbf{e}_{1}}$.
Measuring in a basis that does not resolve the displacement erases this information and restores
the coherence, as in a quantum eraser \cite{Scully82,Walborn2002eraser}. In the far field, for example, the displacement is transformed into a relative phase, and 
entanglement is limited only by the variation of the Bell phase across the apertures. However, the far field
is conjugate to the near field, so a QKD protocol alternating between the two would lose the
polarization channel in half of its rounds. 

In the intermediate planes, on the other hand, each detected coordinate combines position and momentum, and the
displacement of Eq.~(\ref{shiftx}) is partially resolved. The FRFT
divides it between a relative shift of the coincidence patterns and a
gradient of the relative phase of the polarization state \cite{Li2026}, and the coherence
that survives is set by the extent of the detection apertures and the field of view of the collection optics \cite{companion}. For the parameters of our source and a
collection extent comparable to the size of the beams, the concurrence is predicted to be
approximately $0.93$ in the opposite-basis configurations and $0.96$ in the same-basis ones,
compared with zero in the crystal plane. The resource is thus a property of the source together
with the detection system, as is the effective dimension of the spatial entanglement itself
\cite{vanExter2006,Pors2008}.

Selecting two narrow detection regions in each arm then yields a
two-dimensional spatial subspace, and the state in every combination of spatial bases is
approximately
\begin{equation}
|\psi\rangle \propto \left(|\bm{\rho}_{0},\bm{\rho}_{0}\rangle+|\bm{\rho}_{1},\bm{\rho}_{1}\rangle\right)\otimes
\left(|\mathbf{e}_{0},\mathbf{e}_{0}\rangle+|\mathbf{e}_{1},\mathbf{e}_{1}\rangle\right).
\label{III}
\end{equation}
The product form is what allows the two degrees of freedom to be used as independent channels.
It is an approximation whose validity depends on the detection system as well as on the source,
and we test its consequences directly in Sec.~\ref{sec:exp}.

\subsection{Testing the independence of the two channels}
\label{sec:independence}

The polarization state recorded in any configuration has the form described above. With
linear analyzers at angles $\theta_{A}$ and $\theta_{B}$, the correlation function is
\begin{equation}
E(\theta_{A},\theta_{B})=\frac{N_{\uparrow\uparrow}+N_{\downarrow\downarrow}-N_{\uparrow\downarrow}-N_{\downarrow\uparrow}}
{N_{\uparrow\uparrow}+N_{\downarrow\downarrow}+N_{\uparrow\downarrow}+N_{\downarrow\uparrow}},
\label{Ecorr}
\end{equation}
with the arrows labeling the two outputs of each analyzer. The correlation visibilities are its
values in the $\{H,V\}$ and $\{D,A\}$ bases, $V_{HV}=E(0,0)$ and $V_{DA}=E(\pi/4,\pi/4)$, and for
the state of Eq.~(\ref{I}) at any other angles
\begin{equation}
E(\theta_{A},\theta_{B})=V_{HV}\cos2\theta_{A}\cos2\theta_{B}+V_{DA}\sin2\theta_{A}\sin2\theta_{B}.
\label{Evis}
\end{equation}
The CHSH parameter \cite{Clauser69},
$S=|E(\theta_{A},\theta_{B})-E(\theta_{A},\theta'_{B})+E(\theta'_{A},\theta_{B})+E(\theta'_{A},\theta'_{B})|$,
satisfies $S\leq2$ for any local model, and at the usual settings $\theta_{A}=0$,
$\theta'_{A}=\pi/4$, $\theta_{B}=\pi/8$, $\theta'_{B}=3\pi/8$ it equals
$S=\sqrt{2}\,(V_{HV}+V_{DA})$.
An unpolarized background of weight $f$ reduces both visibilities, so that
\begin{equation}
V_{HV}=1-f,
\qquad
V_{DA}=(1-f)\,\mathcal{C},
\label{vis}
\end{equation}
with the relative phase between the two polarization components compensated as described in
Sec.~\ref{sec:exp}. Without compensation $V_{DA}\leq(1-f)\,\mathcal{C}$.
Linear analyzers do not determine $\mathcal{C}$, but they bound
it without any assumption about the form of the state,
\begin{equation}
\mathcal{C}\geq V_{HV}+|V_{DA}|-1 ,
\label{Cbound}
\end{equation}
which follows from the fidelity of the state with the nearest Bell state
\cite{VerstraeteVerschelde02}. The ratio $V_{DA}/V_{HV}=\mathcal{C}$ is free of the
background, and is the quantity we compare between configurations recorded under different
alignment conditions.

If the state has the product form of Eq.~(\ref{III}), three properties follow, each accessible
from the same coincidence data. The polarization error rate must be the same in every combination
of spatial bases and at every pair of spatial cells, the spatial error rate must be the same for
every setting of the polarization analyzers, and the errors in the two degrees of freedom must be
independent. A background that is random
in both degrees of freedom produces errors in both at once, even when the source itself is a
product. For signal error rates $e_{p}$ and $e_{s}$ in the polarization and spatial sectors, and a
background of weight $f_{b}$ distributed equally over the polarization outcomes and the spatial
cells, the measured error rates are
\begin{align}
Qber^{DV}&=(1-f_{b})\,e_{p}+f_{b}/2, \nonumber\\
Qber^{CV}&=(1-f_{b})\,e_{s}+f_{b}/2, \label{noise}\\
Qber^{both}&=(1-f_{b})\,e_{p}e_{s}+f_{b}/4, \nonumber
\end{align}
where $Qber^{both}$ is the rate at which both are in error, Eq.~(\ref{Qbothdef}). The background also
raises the polarization error rate toward $1/2$ in the cells that receive few coincidences from the
source, which are those in which the spatial outcome is wrong. Since the background is common to
all configurations, the third property can be tested by requiring Eq.~(\ref{noise}) to hold with a
single $f_{b}$ for all combinations of bases. Errors due to the which-crystal coupling, on the other
hand, can only appear in the conjugate polarization basis, since the outcomes in the $\{H,V\}$ basis
do not depend on the coherence between the two crystal contributions. We note that independence is
a property of the source rather than a requirement for security, since the key rates of the two
channels can be bounded jointly for any error statistics, Sec.~\ref{sec:keyrate}.

\section{Experimental Implementation}
\label{sec:exp}

\begin{figure}
    \centering
    \includegraphics[width=0.99\linewidth]{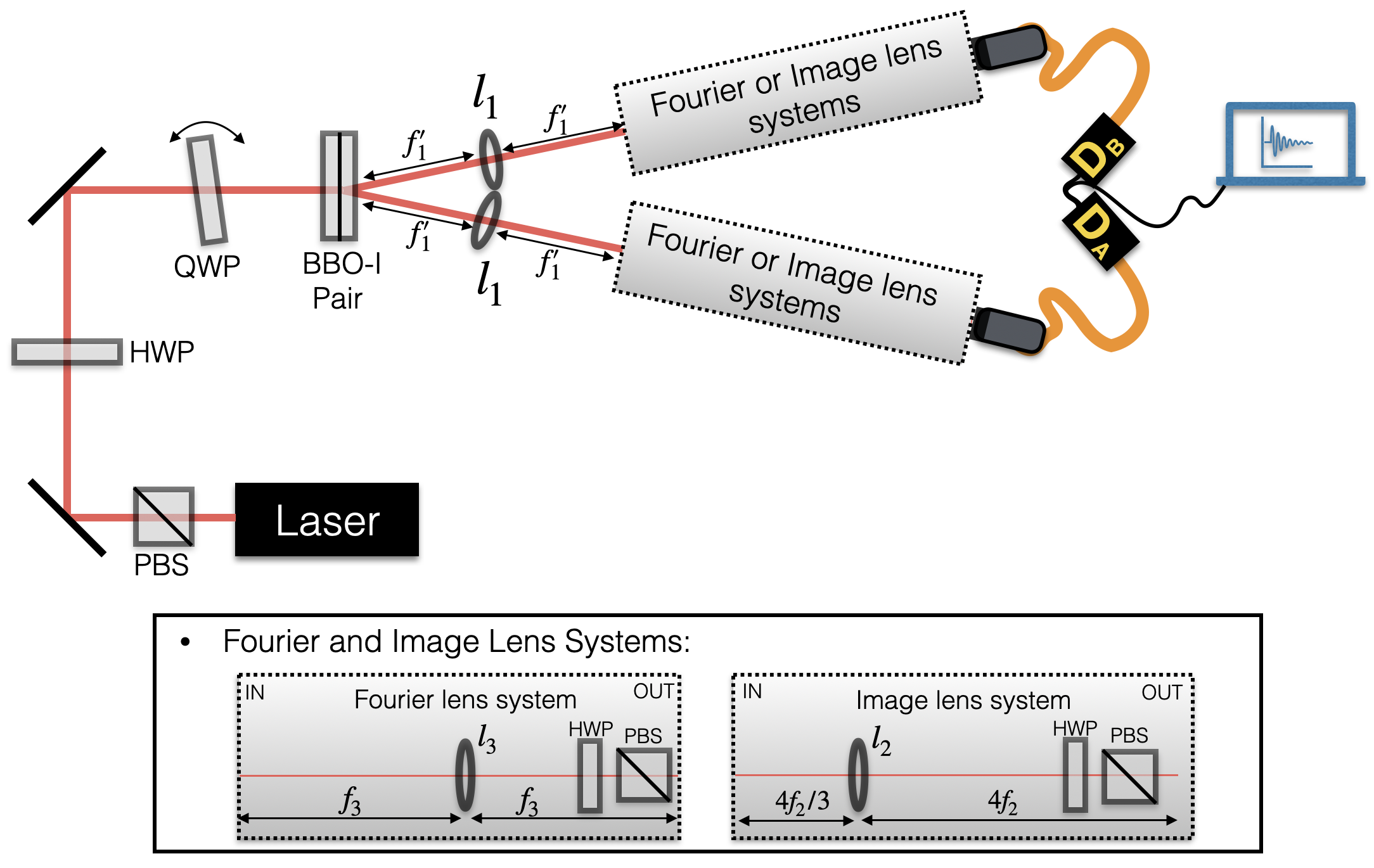}
    \caption{Experimental Setup. See details in the main text. }
    \label{ExpSetup}
\end{figure}

This section describes the implementation of the hyperentangled state at the intermediate plane of the fractional Fourier transform. The experiment is based on spontaneous parametric down-conversion (SPDC), which can simultaneously generate polarization and transverse spatial entanglement. As shown in Fig. \ref{ExpSetup}, we will use SPDC to produce the quantum state of Eq. \eqref{III}. To do so, a continuous-wave diode laser (CrystaLaser, DL-405) operating at $50\,\text{mW}$ generates a collimated beam with a wavelength of $405\,\text{nm}$ and a diameter of $1.2\pm0.2$~mm. This beam is directed through  a polarizing beam splitter (PBS) for initial state preparation, setting the pump polarization state to $\mathbf{e}_H$. Subsequently, another HWP oriented at $22.5^\circ$ rotates the pump polarization to the diagonal direction $\mathbf{e}_D$.

The pump beam illuminates two adjacent 1 mm thick Type I $\beta$-Barium Borate (BBO-I) crystals oriented at $90{\degree}$ with respect to each other, that is, the optical axis of the first lies in the vertical plane, while that of the other lies in the horizontal plane \cite{Kwiat99}. Ideally,  combination of the down-converted photons produced from each crystal would already produce nearly pure Bell states with equal probability amplitudes. In practice, however, birefringence in the nonlinear crystals  introduces a relative phase between the two polarization components. This phase is compensated by slightly tilting a quarter-wave plate (QWP) about the $y$-axis, thereby maximizing the fidelity of the generated Bell state \cite{Kwiat99}.  The generated photons are subsequently directed through independent polarization-analysis and transverse-spatial-analysis stages, allowing the simultaneous characterization of both degrees of freedom. The down-converted
photons are collected in the horizontal plane at an opening angle of approximately $3^{\circ}$
outside the crystals, and the slits used for the spatial measurements are oriented horizontally,
so that they resolve the vertical coordinate and integrate along the emission plane.
Coincidence measurement is performed by collecting photon pairs at 810 nm using interference filters centered at 810 nm with 10 nm bandwidth, followed by a short focal lens which focuses the signal into the multimode optical fiber. The photon pairs are detected by PicoQuant single-photon avalanche diodes (SPADs). The resulting electrical pulses are processed by a Tausand Abacus coincidence-counting module with a $2$~ns coincidence window to record the correlation counts. 

\subsection{Near-field polarization measurements}
 To experimentally investigate the polarization state in the (near field) crystal plane, we performed coincidence measurements with linear-polarization projections in the \{H,V\} and \{D,A\} bases by imaging the crystal plane onto the detection plane with a nominal magnification of $M=-2.61$. The setup consisted of a single lens with a focal length of $f=300$ mm, positioned 415 mm from the crystal and 1080 mm from the detection plane, together with HWPs and PBSs for the linear-polarization projections.

We measured the coincidence rates obtained for joint polarization measurements in the \{H,V\} and \{D,A\} bases when imaging the crystal plane onto the detection plane.

Expressed as correlation visibilities,
these counts give $V_{HV}=0.961\pm0.005$ and $V_{DA}=-0.064\pm0.020$,
and three further runs in the same configuration gave $V_{HV}$ between $0.955$ and $0.977$ and
$|V_{DA}|$ between $0.08$ and $0.12$. This is the behavior predicted in Sec.~\ref{sec:coupling}.
The bound of
Eq.~(\ref{Cbound}) gives $\mathcal{C}\geq0.03\pm0.02$, which does not certify entanglement, and
the corresponding Bell parameter at the usual settings (see below) is $S=1.45\pm0.03$.  The residual $|V_{DA}|$ is
consistent with the finite resolution of the imaging system, since the coherence recovers
rapidly as the detection plane moves away from the crystal plane \cite{companion}.

\subsection{Intermediate plane measurements}
As shown in Fig. \ref{ExpSetup}, the intermediate planes are defined by a single lens $l_1$ with $f_1 = 1000$~mm placed
$300$~mm from the crystal, the plane of interest lies $300$~mm beyond it. The ray-transfer
ABCD matrix from the crystal to that plane is
\begin{equation}
\begin{pmatrix} 0.70 & 510\,\text{mm} \\ -0.001\,\text{mm}^{-1} & 0.70 \end{pmatrix},
\end{equation}
whose equal diagonal elements identify it as a FRFT of order
$\alpha = \arccos(0.70) = 45.6^{\circ}$, i.e.\ $\alpha \simeq \pi/4$, with characteristic
length $\gamma = \sqrt{B/k_{0}\sin\alpha} = 0.303$~mm.

The $r$
basis is then obtained by imaging the  $\alpha \simeq \pi/4$ plane onto the detector with lens $l_2$ ($f_2 = 150$~mm) placed
$500$~mm after $l_1$ and $600$~mm before the detector, giving magnification $M = -3$. The
$s$ basis is obtained by placing lens $l_3$ ($f_3 = 400$~mm) $700$~mm after $l_1$ with the
detector in its focal plane, which measures the transverse momentum at the intermediate
plane, corresponding to $\alpha \simeq 3\pi/4$. The detector coordinates are converted to the dimensionless fractional
variables through Eq.~(\ref{ad}). Slits of widths $\mathcal{S}_1 = 0.25$~mm and
$\mathcal{S}_2 = 0.60$~mm were used for the $\{|s\rangle\}$ and $\{|r\rangle\}$ bases,
respectively.

\begin{figure*}[t]
    \centering
    \makebox[\textwidth][c]{%
\includegraphics[width=0.486\linewidth]{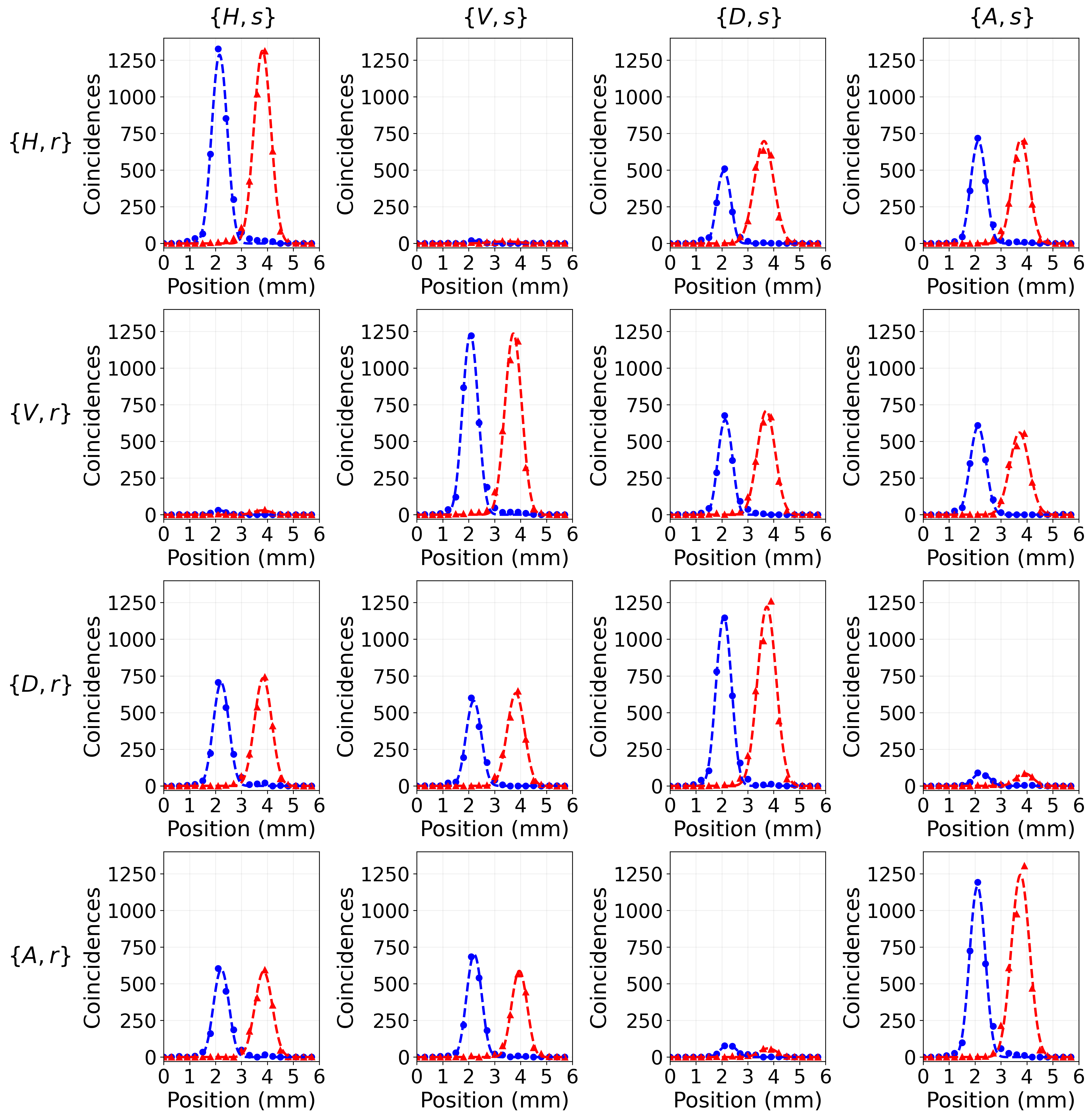}
    \hfill
\includegraphics[width=0.486\linewidth]{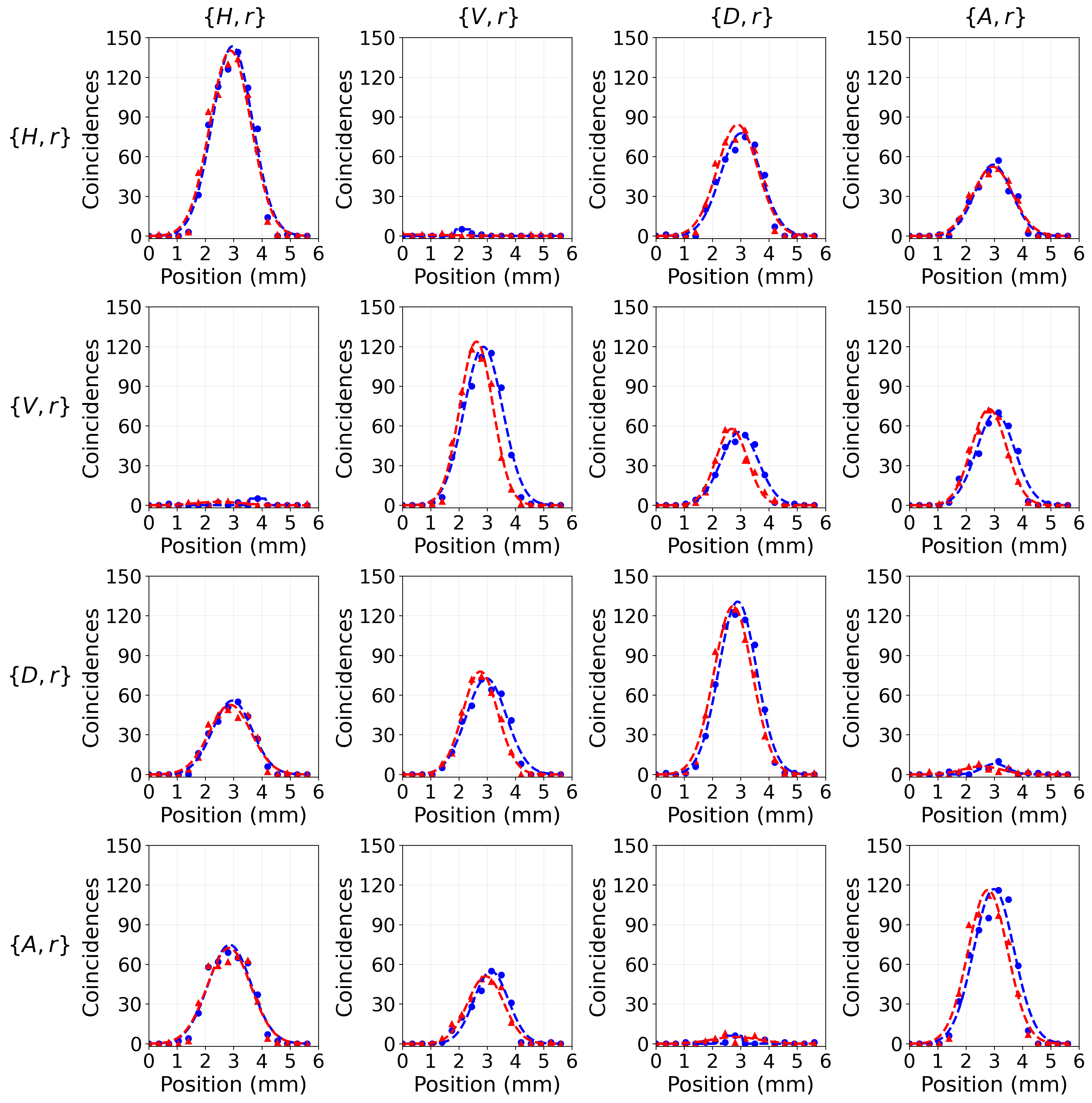}
}
   \vspace{0.1cm}
    
    \makebox[\textwidth][c]{%
\includegraphics[width=0.486\linewidth]{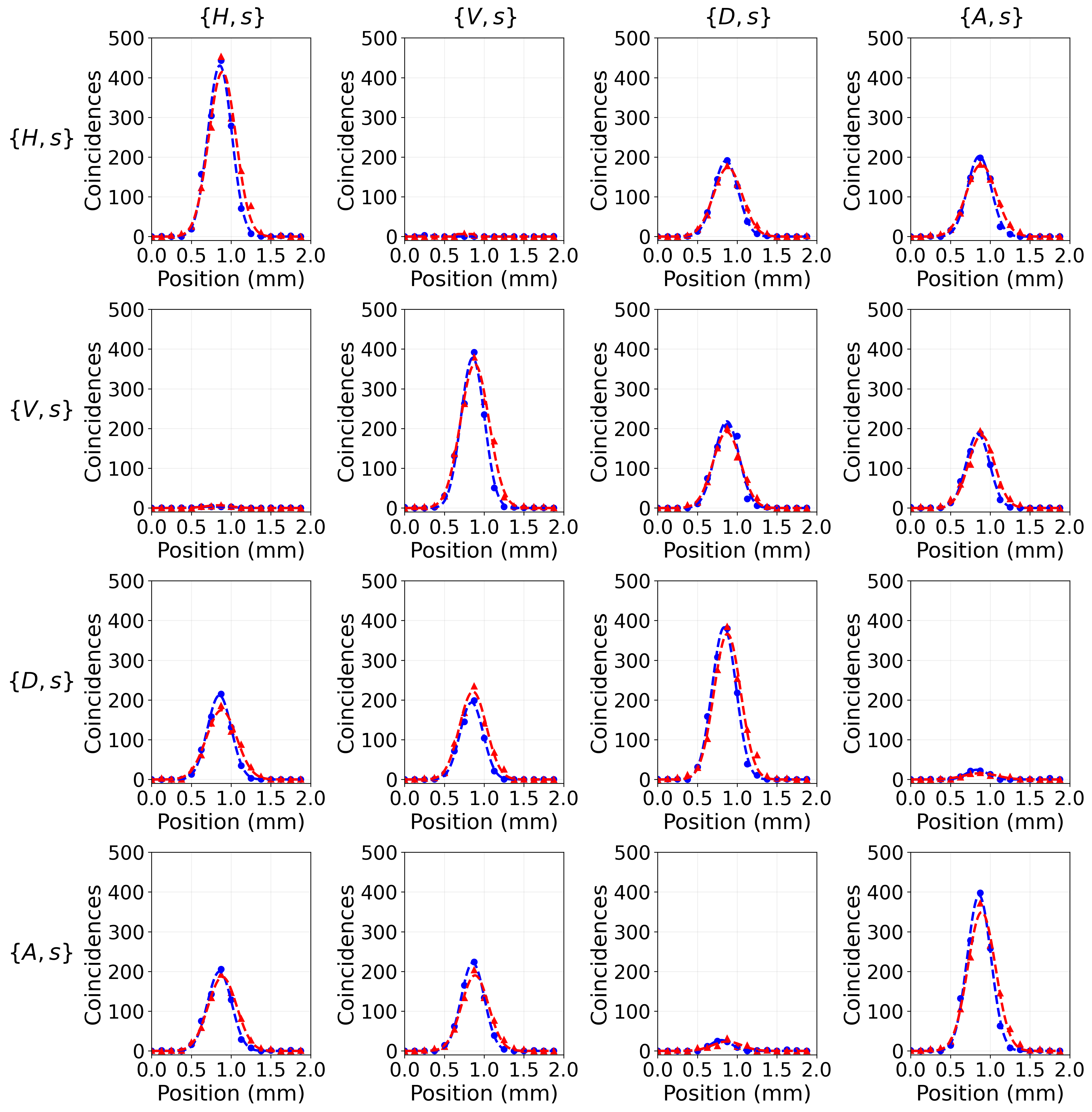}
    \hfill
\includegraphics[width=0.486\linewidth]{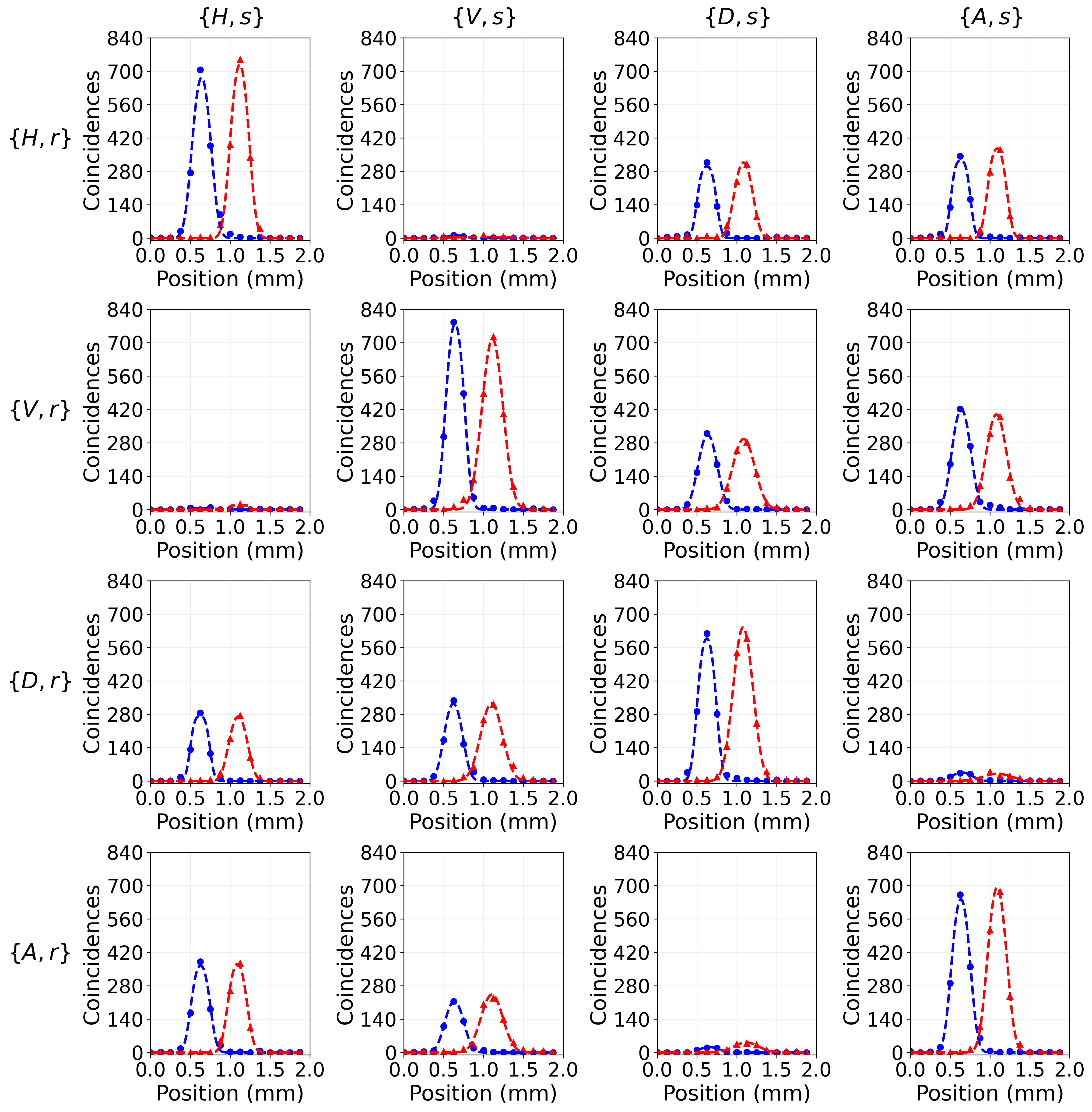}
}
    \caption{Experimental coincidence measurements for the hyperentangled state. Each of the four panels contains a set of coincidence distributions obtained for different combinations of polarization and continuous variable measurement bases. The vertical axis of each subpanel represents the coincidence counts accumulated over 10 seconds [Coincidences], while the horizontal axis shows the transverse position of the signal photon as detector A is scanned [Position (mm)]. The idler detector remains fixed at two transverse positions, denoted by $B_1$ (blue curves) and $B_2$ (red curves). The labels along the rows and columns identify the polarization bases selected by Alice and Bob.}
\label{Resultados}
\end{figure*}

Fig.~\ref{Resultados} presents the coincidence distributions measured for all four combinations $r,s$ measurements conditioned on linear-polarization projections in the \{H,V\} and \{D,A\} bases. The upper-left ($s,r$) and lower-right ($r,s$) panels exhibit the expected displacement of the coincidence peaks as detector A is scanned while detector B remains fixed at positions $B_1$ and $B_2$. Conversely, the upper-right ($r,r$) and lower-left ($s,s$) panels display the overlap predicted for complementary measurement configurations. In addition, the coincidence amplitudes vary according to the selected polarization settings, reflecting the polarization correlations encoded in the Bell state. A quantitative analysis of these measurements is presented in the following subsections. 

\subsection{Certification of the polarization sector}
\label{sec:certif}

We measured the joint visibilities $V_{HV}$ and $V_{DA}$ as well as the CHSH parameter of Sec.~\ref{sec:independence} at the usual polarization settings.
Table~\ref{tab:S} collects the polarization measurements in the four combinations of
intermediate bases, together with the crystal plane for comparison. In all four intermediate
configurations the Bell parameter exceeds the local bound by at least seven standard deviations,
and Eq.~(\ref{Cbound}) certifies a concurrence of at least $0.81$. The ratio $V_{DA}/V_{HV}$,
which is free of background, lies between $0.91$ and $1.01$ in the four configurations, while in
the crystal plane it is $0.07$. The polarization channel is therefore available in every
combination of the two spatial bases, which is the first of the properties of
Sec.~\ref{sec:independence}. The variation between configurations, of $0.1$ in $S$, is comparable to that between repeated alignments of the same
configuration. 

\begin{table}
\centering
\caption{Polarization correlation visibilities and CHSH parameter in the four combinations
of intermediate bases, with the crystal plane for comparison. Uncertainties are statistical. The
last column is the lower bound of Eq.~(\ref{Cbound}) on the concurrence. In the crystal plane $S$
is obtained from the visibilities. }
\label{tab:S}
\begin{tabular}{lcccc}
\hline\hline
Configuration & $V_{HV}$ & $V_{DA}$ & $S$ & $\mathcal{C}\geq$ \\
\hline
$\{s^{A},s^{B}\}$ & $0.933(8)$  & $0.882(11)$  & $2.570\pm0.034$ & $0.81$ \\
$\{s^{A},r^{B}\}$ & $0.911(13)$ & $0.921(12)$  & $2.684\pm0.043$ & $0.83$ \\
$\{r^{A},r^{B}\}$ & $1.000(4)$  & $0.912(24)$  & $2.65\pm0.09$   & $0.91$ \\
$\{r^{A},s^{B}\}$ & $0.944(7)$  & $0.890(10)$  & $2.600\pm0.032$ & $0.83$ \\
Crystal plane     & $0.961(5)$  & $-0.064(20)$ & $1.45\pm0.03$   & $0.03$ \\
\hline\hline
\end{tabular}
\end{table}

\subsection{CV EPR Inequalities}
\label{sec:CVEPR}
\begin{table}
\centering
\caption{Conditional distribution widths $\Delta(A|B)$ for the different combinations of measurement bases, obtained from Gaussian fits to the measured distributions, with the width of the scanned slit
deconvolved and that of the fixed slit included, together with the corresponding EPR-steering parameter $S^{CV}$. The
bound for a non-steerable state is $S^{CV} \geq 1/4$.}
\label{tab:Reid}
\begin{tabular}{lccc}
\hline\hline
Basis combination & $\Delta(A|B)$& $S^{CV}$& Steering \\
\hline
$\{r^{A},s^{B}\}$& $0.509\pm0.013$ &  &  \\
$\{s^{A},r^{B}\}$& $0.525\pm0.028$ & $0.071 \pm 0.009$ & yes \\\hline
$\{r^{A},r^{B}\}$ & $0.868\pm0.009$
\\
$\{s^{A},s^{B}\}$& $0.875\pm0.011$& $0.577 \pm 0.086$   & no  \\
\hline\hline
\end{tabular}
\end{table}

To benchmark the correlations between the transverse variables in the different measurement bases, we measure the conditional distributions and investigate the presence of EPR steering by testing the conditional-variance EPR steering inequality \cite{Reid89}:
\begin{equation}
S^{CV} =  \Delta^2(r^A_i|s^B_k)\Delta^2(s^A_j|r^B_k) \geq \frac{1}{4}, \label{EPR}
\end{equation}
where the subscripts $i$ and $j$ denote the scanned positions of the
variables $r$ and $s$, respectively, at Alice's detector (signal),
conditioned on the fixed positions of Bob's detector (idler), indexed
by $k=1,2$. Note that the conditional variances are evaluated at only
two fixed positions of Bob's detector, rather than averaged over his
full distribution. If the product of the conditional variances
violates the inequality in Eq.~\eqref{EPR}, EPR steering is certified.

We define the dimensionless fractional position $r$ and momentum $s$ of the intermediate plane as \cite{PhysRevA.79.033801},

\begin{equation}
r = \frac{1}{M\gamma}\rho'_{det}, \qquad  s  = \gamma\frac{k_0}{f_3}\rho''_{det},
\label{ad}
\end{equation}
where $\rho'_{det}(\rho''_{det})$ denotes the position in the detector's transverse plane, while $1/M\gamma=1.10$~mm$^{-1}$ and $\gamma k_{0}/f_{3}=5.876$~mm$^{-1}$.
The measurements in the $\{r,s\}$ and $\{s,r\}$ bases exhibited quantum correlations that violated the EPR-steering criterion, whereas measurements in the $\{r,r\}$ and $\{s,s\}$ bases showed no evidence of EPR steering.
The conditional widths in the two opposite-basis configurations agree within their
uncertainties, as expected from Eq.~(\ref{intvar}), since both are governed by the same pair of
narrow source variables.

Steering criteria can also be cast in terms of entropies \cite{Walborn11,Schneeloch2013}, and
have been tested with spatially entangled photons \cite{Dixon2012,Howland2013}.
The same entropies bound the secret key available from an encoding in the continuous variables
themselves, in a setting that is device independent on one side
\cite{Branciard2012,Furrer2012,Gehring2015}, and they have been used to quantify the secure
information capacity of spatially entangled photons \cite{Almeida2005,Leach2012}. The key per
detected pair satisfies
\begin{equation}
K\geq\log_{2}\kappa-h(r^{A}|s^{B})-h(s^{A}|r^{B}),
\label{Kent}
\end{equation}
with the entropies in bits, where the constant $\kappa$ comes from the uncertainty relation for
Alice's two conjugate measurements. If the eavesdropper's information is classical,
$\kappa=\pi e$, the constant of the entropic steering criterion \cite{Walborn11}. If she holds a
quantum memory, $\kappa=2\pi$ \cite{Berta2010,Furrer2014}, which is the case relevant for
security.  A Gaussian
distribution has the largest entropy for a given variance, so
$h(r^{A}|s^{B})\leq\tfrac{1}{2}\log_{2}[2\pi e\,\Delta^{2}(r^{A}|s^{B})]$, and similarly for the
other term. In terms of the conditional variances of Eq.~(\ref{EPR}), Eq.~(\ref{Kent}) then gives
\begin{equation}
K\geq-\tfrac{1}{2}\log_{2}\!\left[\left(\frac{2\pi e}{\kappa}\right)^{2}S^{CV}\right].
\label{Kgauss}
\end{equation}
For $\kappa=\pi e$ this is positive exactly when the Reid criterion is violated, $S^{CV}<1/4$,
while for $\kappa=2\pi$ it requires $S^{CV}<1/e^{2}\simeq0.135$. The measured $S^{CV}=0.071$ of
Table~\ref{tab:Reid} gives $K\geq0.91$ bits per detected pair against a classical adversary and
$K\geq0.47$ bits against a quantum one.

This is well below what the source could provide. For the double-Gaussian state the conditional
variances follow from Eq.~(\ref{condvar}). In the near and far fields
$\Delta^{2}(\rho^{A}|\rho^{B})=1/2(\sigma_{+}^{2}+\sigma_{-}^{2})$ and
$\Delta^{2}(q^{A}|q^{B})=2\sigma_{+}^{2}\sigma_{-}^{2}/(\sigma_{+}^{2}+\sigma_{-}^{2})$, so that
Eq.~(\ref{Kgauss}) gives
\begin{equation}
K_{nf}=\log_{2}\frac{\kappa(\sigma_{+}^{2}+\sigma_{-}^{2})}{2 \pi e\,\sigma_{+}\sigma_{-}}\simeq
\log_{2}\frac{\kappa \sigma_{-}}{2 \pi e\,\sigma_{+}} .
\label{Knf}
\end{equation}
In the opposite-basis intermediate configurations both conditional variances are given by
Eq.~(\ref{condopp}), and
\begin{equation}
K_{int}\simeq-\log_{2}\left[\frac{\kappa}{2\pi e}\left(\sigma_{+}^{2}+\frac{1}{4\sigma_{-}^{2}}\right)\right].
\label{Kint}
\end{equation}
The two differ by
\begin{equation}
K_{nf}-K_{int}\simeq\log_{2}\left(\sigma_{+}\sigma_{-}+\frac{1}{4\sigma_{+}\sigma_{-}}\right)\geq0,
\label{price}
\end{equation}
which is the price of using the intermediate bases. It vanishes when the near-field and
far-field correlation widths are equal, $2\sigma_{+}\sigma_{-}=1$, and grows with the imbalance
between them, since the intermediate bases average the two widths.

The two source parameters follow from the pump and the crystal. For a pump of $1/e^{2}$ radius
$w_{p}$, $\sigma_{+}^{2}=\gamma^{2}/2w_{p}^{2}$, and the Gaussian approximation to the phase
matching gives $\sigma_{-}^{2}=k\gamma^{2}/0.455\,\ell$ \cite{Chan2007}, with $k$ the
down-converted wave number in the crystal. In the scanned direction, where the phase matching has
the collinear form, $w_{p}=0.6\pm0.1$~mm, $\ell=1$~mm and $\gamma=0.303$~mm give
$\sigma_{+}^{2}\simeq0.13$ and $\sigma_{-}\simeq51$, so that $K_{nf}\simeq5.7$ bits and
$K_{int}\simeq1.5$ bits, both above the measured value. The intermediate value is set almost
entirely by the pump, $K_{int}\simeq\log_{2}(2w_{p}^{2}/e\gamma^{2})$, so that each doubling of the
pump waist would add two bits. The near-field advantage, however,
requires resolving the near-field correlation width, $\gamma/2\sigma_{-}\simeq3~\mu$m at the
crystal, far below the resolution of practical imaging systems.

The widths of Table~\ref{tab:Reid} describe detector A at full resolution, given the actual
resolution of detector B, since the scanned slit is deconvolved and the fixed slit is included.
The comparison with the source is therefore with the intrinsic value
$\sigma_{+}^{2}+1/4\sigma_{-}^{2}\simeq0.13$ plus the variance of the fixed slit, which in
dimensionless units is $0.18$ for the $0.25$~mm slit of the $s$ readout and $0.036$ for the
$0.6$~mm slit of the $r$ readout. For $\{r^{A},s^{B}\}$ this gives about $0.31$, against the
measured $0.26$, in agreement within the uncertainty of the pump diameter. For $\{s^{A},r^{B}\}$ it
gives about $0.17$, against the measured $0.28$. The different-basis conditional variance is also
sensitive to small deviations of the fractional orders from $\alpha+\beta=\pi$, which admit a
fraction of the broad global variance and may account for the remaining difference. The fixed
slit of the $s$ readout is the largest single contribution, and a narrower slit there, or a
longer focal length $f_{3}$, would bring the rate closer to the intrinsic value of about $1.5$
bits for the intermediate bases.

\subsection{Quantum Bit Error Rate}

While the violations of the CHSH and EPR inequalities presented in the previous subsections certify the presence of quantum correlations in the polarization and continuous-variable subsystems, respectively, the practical performance of hyperentanglement in a QKD protocol is ultimately characterized by its quantum bit error rate (QBER). This quantity measures the fraction of erroneous bits shared by Alice and Bob and determines whether a secure secret key can be distilled after the classical stages of error correction and privacy amplification.

The QBER can be defined as
\begin{equation}
    Qber = \frac{P_{wrong}}{P_{wrong}+P_{right}}
\end{equation}
where $P_{\mathrm{wrong}}$ and $P_{\mathrm{right}}$ denote the probabilities of obtaining incorrect and correct symbols, respectively. Throughout this analysis, the QBER is evaluated only for events in which Alice and Bob choose compatible measurement bases, since only these events contribute to the sifted key. Consequently, low QBER values indicate both high-fidelity state preparation and the feasibility of secure key generation.

Let
$P^{ji}_{mn}$ denote the joint probability that Alice records polarization outcome $j$ at
transverse position $m$ and Bob records polarization outcome $i$ at position $n$, in a given
pair of compatible bases. The labels are assigned so that a correlated event has $j=i$ and
$m=n$. The three error rates of interest are then
\begin{align}
Qber^{DV} &= \sum_{j\neq i}\sum_{m,n} P^{ji}_{mn}\,, \label{QDVdef}\\
Qber^{CV} &= \sum_{j,i}\sum_{m\neq n} P^{ji}_{mn}\,, \label{QCVdef}
\end{align}
with $\sum_{j,i}\sum_{m,n}P^{ji}_{mn}=1$. Equations
(\ref{QDVdef}) and (\ref{QCVdef}) are the error rates of the two sectors taken separately,
obtained by marginalizing over the other sector. The probability that both sectors are in
error is
\begin{equation}
Qber^{both}=\sum_{j\neq i}\sum_{m\neq n}P^{ji}_{mn} ,
\label{Qbothdef}
\end{equation}
which for independent errors and no background equals $Qber^{DV}\,Qber^{CV}$, and in general is
given by Eq.~(\ref{noise}).

\begin{table}[htbp]
\centering
\footnotesize
\caption{Error rates in percent in the two configurations of opposite spatial bases, pooled
over the four pairs of spatial cells. $N$ is the number of coincidences, $Qber^{both}$ is the rate
at which both sectors are in error, Eq.~(\ref{Qbothdef}), and the last column is the product
$Qber^{DV}Qber^{CV}$ expected for independent errors without a common background. Uncertainties are statistical,
$[Q(1-Q)/N]^{1/2}$.}
\label{tab:Qber}
\begin{tabular}{llccccc}
\hline\hline
Spatial & Pol. & $N$ & $Qber^{DV}$ & $Qber^{CV}$ & $Qber^{both}$ & Product \\
\hline
$\{r^{A},s^{B}\}$ & $\{H,V\}$ & 5217 & 2.15(20) &  1.42(16) & 0.23(7) & 0.03 \\
$\{r^{A},s^{B}\}$ & $\{D,A\}$ & 5252 & 6.00(33) & 0.86(13) & 0.19(6) & 0.05 \\
$\{s^{A},r^{B}\}$ & $\{H,V\}$ & 3038 & 1.45(22) & 1.09(19) & 0.16(7) & 0.02 \\
$\{s^{A},r^{B}\}$ & $\{D,A\}$ & 2687 & 4.50(40) & 0.67(16) & 0.22(9) & 0.03 \\
\hline\hline
\end{tabular}
\end{table}

\begin{table}[htbp]
\centering
\footnotesize
\caption{$Qber^{DV}$ in percent for each pair of spatial cells, in the two polarization
bases. In the opposite-basis configurations the cells with different indices are spatial errors,
which have few counts. Upper limits are one-sided at $84\%$ confidence, $1.84/N$.}
\label{tab:QberCells}
\begin{tabular}{lcc|lcc}
\hline\hline
Cells & $\{H,V\}$ & $\{D,A\}$ & Cells & $\{H,V\}$ & $\{D,A\}$ \\
\hline
$\{r_1,s_1\}$ & 1.96(27) & 6.67(50) & $\{s_1,r_1\}$ & 0.80(23) & 3.97(54) \\
$\{r_2,s_2\}$ & 1.93(27) & 5.11(42) & $\{s_2,r_2\}$ & 1.80(34) & 4.64(58) \\
$\{r_1,s_2\}$ & 32(8)    & 27(11)   & $\{s_1,r_2\}$ & 16(8)    & 25(15)   \\
$\{r_2,s_1\}$ & $<5$     & 20(7)    & $\{s_2,r_1\}$ & 14(9)    & 40(15)   \\
\hline
$\{r_1,r_1\}$ & $<2.7$   & 3.2(2.2) & $\{s_1,s_1\}$ & 1.03(59) & 6.1(1.4) \\
$\{r_2,r_2\}$ & $<2.4$   & 5.6(2.7) & $\{s_2,s_2\}$ & 0.30(30) & 7.2(1.5) \\
$\{r_1,r_2\}$ & 3.1(1.7) & 4.6(2.2) & $\{s_1,s_2\}$ & 1.52(75) & 6.7(1.7) \\
$\{r_2,r_1\}$ & 4.0(1.8) & 2.7(1.5) & $\{s_2,s_1\}$ & $<1.5$   & $<1.8$   \\
\hline\hline
\end{tabular}
\end{table}

Table~\ref{tab:Qber} summarizes the error rates in the two configurations of opposite spatial
bases, which are those that contribute to the sifted key. The polarization error rate is between
$1.5$ and $2.2\%$ in the $\{H,V\}$ basis and between $4.5$ and $6.0\%$ in $\{D,A\}$, and the spatial
error rate is between $0.7$ and $1.4\%$. The joint error rate is larger than the
product of the two, which is the effect of the background described in Sec.~\ref{sec:independence}.
The cells in which the spatial outcome is wrong receive few coincidences from the source, so that
the background dominates the counts. The polarization error rate in these cells is
$\sim 16\%$ in the $\{H,V\}$ basis and $\sim 25\%$ in $\{D,A\}$, against $1.7\%$ and $5.3\%$ in the cells with
the correct spatial outcome, as seen in Table~\ref{tab:QberCells}.

We fit Eq.~(\ref{noise}) to the distributions of the four error patterns in the four rows of
Table~\ref{tab:Qber} by maximum likelihood with Poisson statistics, with independent signal error
rates $e_{p}$ and $e_{s}$ for each row and a single background weight $f_{b}$. The fit gives
$f_{b}=(0.74\pm0.13)\%$, with a deviance of $0.5$ for $3$ degrees of freedom, while the model without
background, $Qber^{both}=Qber^{DV}Qber^{CV}$, is rejected with a deviance of $75$ for $4$ degrees of
freedom. The error statistics are thus those of a source that factorizes between the two degrees of
freedom, together with a background that is random in both, which is the third property of
Sec.~\ref{sec:independence}. The excess joint error rate is present in the $\{H,V\}$ basis, in which
the which-crystal coupling cannot produce errors, as well as in $\{D,A\}$. The background adds about
$0.37\%$ to each error rate, roughly a third of the spatial error rate, and the fitted signal error
rates are $e_{p}$ between $1.1$ and $1.8\%$ in $\{H,V\}$ and between $4.1$ and $5.7\%$ in $\{D,A\}$,
and $e_{s}$ between $0.3$ and $1.0\%$.

Table~\ref{tab:QberCells} tests the first property of Sec.~\ref{sec:independence} at the level
of individual cells. In the cells with the correct spatial outcome, and in all cells of the
same-basis configurations, $Qber^{DV}$ is at most $4\%$ in the $\{H,V\}$ basis and at most $7\%$
in $\{D,A\}$, so that the polarization channel is available in every combination of spatial bases.
Within each configuration the error rates of the cells agree at about the $5\%$ significance level,
with $\chi^{2}=15.5$ and $15.2$ for $8$ degrees of freedom in the $\{H,V\}$ and $\{D,A\}$ bases. The
differences between runs are larger, as seen by comparing with the visibilities of Table~\ref{tab:S}, for which
$Qber^{DV}=(1-V)/2$.

The spatial error rate was also recorded for all sixteen combinations of the polarization
analyzers in the $\{H,V\}$ and $\{D,A\}$ bases, including those in which Alice and Bob use different
polarization bases. For the twelve combinations without a polarization error it lies between $0$ and
$1.8\%$ in the $\{r^{A},s^{B}\}$ configuration and between $0.1$ and $2.2\%$ in $\{s^{A},r^{B}\}$. The
count-weighted averages are $0.97(10)\%$ and $0.72(11)\%$ when the polarization bases agree, and
$0.75(9)\%$ and $1.20(15)\%$ when they differ. The spatial channel is thus available for every setting
of the polarization analyzers, which is the second property of Sec.~\ref{sec:independence}. 

\subsection{Asymptotic secret key rates}
\label{sec:keyrate}

The error rates measured above determine the asymptotic secret key rate that the source would
support. For an entanglement-based protocol with one-way postprocessing, the rate per sifted pair in
each sector is bounded by \cite{Shor00,Gisin02}
\begin{equation}
R \geq 1 - h(Q_{z}) - h(Q_{x}),
\end{equation}
where $h$ is the binary entropy and $Q_{z},Q_{x}$ are the error rates in the two conjugate bases.
Pooling the two configurations of Table~\ref{tab:Qber}, the polarization sector has $Q_{z}=1.89\%$
and $Q_{x}=5.49\%$, giving $R^{DV}\geq0.558\pm0.013$ secret bits per sifted pair, and the spatial
sector has $Q_{z}=1.14\%$ and $Q_{x}=0.89\%$, giving $R^{CV}\geq0.837\pm0.011$, with uncertainties
obtained by resampling the coincidence counts. A pair for which both bases match thus carries at
least $R^{DV}+R^{CV}=1.40\pm0.02$ secret bits, compared with at most one bit for a single qubit
channel and two bits for an ideal two-channel resource. This sum is a lower bound for any error
statistics, since the entropy of the joint errors is at most the sum of the entropies of the
separate errors, so that correlations between them can only increase the rate \cite{Berta2010}.
Here the spatial sector is treated as a qubit encoded in two transverse cells, which is an
idealization, since events outside the two cells are discarded. The continuous-variable bound of
Eq.~(\ref{Kent}) does not rely on this assumption. Increasing the number of transverse cells would
raise $R^{CV}$ toward $\log_{2}d$ for a $d$-cell encoding.

It is important to emphasize that the single-detector scanning measurement is used solely as a characterization method and does not constitute a practical implementation of the protocol. A realistic implementation of the encoding scheme would require multiplexed detection.  Both channels rely on selecting detection events, through the apertures in the spatial
sector and the coincidence window. Restricting the apertures is a local filtering operation, which
can increase the entanglement of the selected events but cannot create it, and in an
entanglement-based protocol it does not by itself compromise security, provided that the
probability of detection does not depend on the choice of basis or on the outcome. Here the two
spatial bases are read out with different optics and slit widths, so their detection efficiencies
need not be equal. Mismatches of this kind are a known vulnerability, including for spatial modes
\cite{Makarov2006,Fung2009,Sajeed2015}, and a security analysis would have to bound them.

\section{Conclusion}
In this work we have characterized a photon-pair source hyperentangled in polarization and
in the transverse spatial degree of freedom, with the spatial sector measured entirely in
intermediate planes of the fractional Fourier transform. In the non-collinear two-crystal source
the pairs from the two crystals are displaced with respect to one another, and detection in the
crystal plane reads out this which-crystal information completely, as confirmed by the vanishing
conjugate-basis visibility measured there. The polarization entanglement is not destroyed but made
inaccessible, such that measuring the transverse variables in a basis that does not resolve the
displacement erases the which-crystal information. The intermediate planes provide two conjugate
spatial bases in which this erasure is partial, and the polarization entanglement survives in all
four of their combinations.

Applying Bell and EPR criteria to the same photon pairs certifies both sectors simultaneously.
The polarization visibilities and Bell parameters are comparable in every combination of spatial
bases, the spatial error rates remain below $2.5\%$ for every setting of the polarization
analyzers, and the joint error statistics are those of a source that factorizes between the two
degrees of freedom, together with a background of about $0.7\%$ that is random in both.

Expressed in the figures of merit of entanglement-based key distribution, the source
supports asymptotic secret key rates of $0.56$ and $0.84$ bits per sifted pair in the
polarization and spatial sectors respectively, or about $1.4$ bits per pair when both are
used. We emphasize that this is a
characterization of a source rather than a key-distribution experiment: the measurements
were performed with scanning single detectors over a laboratory-scale path, no key was
distributed, and no finite-key analysis has been attempted. What the measurements do
establish is that a single SPDC source, read out in intermediate phase-space planes,
provides two separately certified channels whose independence can be tested directly
from their joint error statistics, which is the resource such protocols require. Although only
two transverse cells were used here, the spatial encoding extends directly to a larger number of
cells, which would increase the key rate per pair.

\section*{Acknowledgements}
This work was supported by the Brazilian funding agencies: Conselho Nacional de Desenvolvimento Cient\'ifico e Tecnol\'ogico (CNPq) [DOI 501100003593, Grant No. 306105/2022-5, Grant No. 312173/2022-9, Grant No. 303145/2026-9], Coordena\c{c}\~ao de Aperfei\c{c}oamento de Pessoal de N\'ivel Superior (CAPES) [DOI 501100002322], Funda\c{c}\~ao de Amparo \`a Pesquisa do Estado de Goi\'as (FAPEG) [Grant No. LAM2026011000041, Grant No. 202510267001843], Funda\c c\~{a}o de Amparo \`{a} Pesquisa do Estado de Santa Catarina (FAPESC)
[DOI 501100005667, DOI 2025TR001683], Instituto Nacional de Ci\^encia e Tecnologia de Infraestruturas Qu\^antica e Nano para Aplica\c c\^oes Convergentes  (406636/2022 2), and Instituto Nacional de Ci\^encia e Tecnologia de Dispositivos Qu\^anticos (408783/2024 9).  SPW was supported by Fondo Nacional de Desarrollo Científico y Tecnológico (FONDECYT) Grant No. 1240746,  ANID Anillo Project ATE250003, and ANID Millennium Science Initiative Program -- ICN17$_-$012.

\bibliographystyle{apsrev4-2}
\bibliography{refs}

\end{document}